# Analyzing Nanogranularity of Focused Electron-Beam-Induced Deposited (FEBID) Materials by Electron Tomography

*Cornelia Trummer[1], Robert Winkler[2], Harald Plank[1,2,3], Gerald Kothleitner[1,3],\*, Georg Haberfehlner[3,\**

[1] Graz Centre for Electron Microscopy, 8010 Graz, Austria.

[2] Christian Doppler Laboratory DEFINE – Institute of Electron Microscopy and Nanoanalysis, Graz University of Technology, 8010 Graz, Austria.

[3] Institute of Electron Microscopy and Nanoanalysis, Graz University of Technology, 8010 Graz, Austria.

ABSTRACT

Nanogranular material systems are promising for a variety of applications in research and development. Their physical properties are often determined by grain sizes, shapes, mutual distances and by the chemistry of the embedding matrix With focused electron beam induced deposition arbitrarily shaped nanocomposite materials can be designed, where metallic, nanogranular structures are embedded in a carbonaceous matrix. Using "post-growth" electron beam curing, these materials can be tuned for improved electric transport or mechanical behavior. Such an optimization necessitates a thorough understanding and characterization of the internal changes in chemistry and morphology, which is where conventional projection based imaging techniques fall short. Here, we apply scanning transmission electron tomography to get a comprehensive picture of the distribution and morphology degree of embedded Pt nanograins after initial fabrication, and we demonstrate the impact of electron beam curing, which leads to condensed regions of interconnected metal nanograins.

KEYWORDS: STEM tomography, nanogranular materials, focused electron beam induced deposition, platinum, nanostructures

In this study, we focus on metal reinforced carbon matrix composite materials, fabricated by focused electron beam induced deposition. (**FEBID**).[1] The electric charge transport within these materials is dominated by tunneling between the metallic grains in the variable range hopping regime.[2,3] Depending on the coupling strength, which is determined by grain diameter, grain distances and by the dielectric constant of the separating matrix, such materials can reveal transport behavior from the insulating to the metallic regime.[4–6] Typically grain diameters are that small, that the intragrain conductance is far greater than the intergrain tunneling coupling.[2] Initial fabrication parameters can set properties in a small window[7], while post-growth electron-beam curing enables tuning of materials properties over a wide range[4,8]. This opens up different application possibilities as recently reviewed by Huth[3,9,10] or Winkler et al.[11]. Figure 1 shows the two principal fabrication steps of deposition and curing, investigated in this work along with transmission electron microscopy (TEM) images. The origin of the nanogranularity for most FEBID materials lies in the fabrication process itself. A gaseous precursor is injected into a scanning electron microscope (**SEM**) via a fine capillary in close proximity to the region, where the material should be deposited. After physisorption onto the surface, a diffusion phase follows until the precursor desorbs again. A focused electron beam is then used for local precursor dissociation in the diffusion phase, leading to fragmentation and immobilization, forming the intended, functional deposit.[11] FEBID typically works with organometallic precursors. Trimethyl(methylcyclopentadienyl)platinum(IV) ($C_5H_4CH_3Pt(CH_3)_3$), as used in this study, yields comparably high carbon contents (up to 90 at.%) in the deposits, the carbon forms an insulating matrix.[7] Within this matrix, fully fragmented metal atoms condense into spatially embedded metal grains of 1.5 – 3 nm in size, depending on the precursor and initial fabrication conditions[7] (for more details see section 1 of the Supporting Information).

As electrical[8,12], mechanical[13,14] and thermal properties[15] of such nanogranular materials depend strongly on the inner composition (grain size, -shape, inter-granular distances and chemical matrix properties), proper material tuning is highly relevant. The latter is mostly done by the aforementioned post-growth electron beam curing (**EBC**), first reported by Porrati et al.[4] and Plank et al.[8] for Pt based materials. Their findings suggested, that EBC leads to Pt grain growth caused by the completion of dissociation processes, which also leads to reduced inter-granular distances. This, in turn, changes the coupling strength due to reduced Coulomb barriers, increasing electric conductivities up to 4 orders of magnitude.

Although consistent with theory and simulations,[2] supporting experimental evidence, regarding the shape of grains, as well as their spatial distribution and possible percolation paths, was unavailable at that time.

In this study, we aim at providing missing information about the internal morphologies of Pt-C nanopillars as-deposited (**AD**) and after EBC using scanning transmission electron microscopy (STEM) based 3D nano-tomography.[16] Full-range tomography is done on pillar shaped samples. In the last few years, 3D-printing of meshed-styled nano-objects via FEBID has undergone significant progress[11], which opened up new possibilities for diverse applications such as plasmonics[17], nano-magnetics[18], gas sensors[8] or nanoprobe concepts[15]. The smallest building blocks of complex geometries needed for these applications are straight nano-pillars, which are in the center of this study. Electron tomography in a scanning transmission electron microscope (**STEM**) is often used for studying shapes, sizes and distribution of e.g. nanoparticles or pores in different materials.[19–24] While high-angle annular dark-field (**HAADF**) imaging[25] clearly distinguishes Pt grains from the carbon matrix, sequential image acquisition at tilt angles from

- 90° to + 90° paves the way for a 3D reconstruction.[25] Thereby, we receive information about the shape, size and spatial distribution of Pt grains within the Pt-C nano-pillars.

Two types of samples have been investigated: *1)* AD and *2)* EBC treated Pt-C FEBID nano-pillars. Two-dimensional (**2D**) HAADF STEM projection images (figure 1) of the pillar top region already reveal different grain sizes before (a) and after EBC (b). However, from these images a determination of grain sizes, -shapes and inter-granular distances is hampered due to their 2D projection nature. Furthermore, it is not possible to decide whether the Pt grains are mutually interconnected, or well separated. Hence, tomographic tilt series were acquired, aligned, reconstructed and segmented to reveal the spatial distribution of Pt grains (see Electron Tomography in section 2 of the Supporting Information). Figure 2 shows tomographic reconstructions for AD (a) and EBC samples (b), taken from the tip region. (see also Figure S3 for cross-sections) First, 3D reconstructions confirm previous results of increasing grain sizes after EBC.[4] However, figure 2b already reveals that grain-shapes have strongly changed and the spatial distribution has become less homogeneous. Some of the grain parameters are summarized in figure 3. The grain volumes (figure 3a) and thus the size of the grains, increase during the EBC treatment. Figure 3b displays the sphericity $\varphi$, which is calculated as the ratio of the surface area of a sphere with the same volume as the particle to the actual surface area of the particle. It illustrates that hardly any spherical (value 1) but mostly elongated grains as well as partially interconnected grains exist, as shown by the inset examples. Interestingly, the distribution is not strongly changed by EBC (compare red and blue bars). Inspection of the grain-to-grain distances in figure 3c, shows that the separation in EBC treated pillars feature less contributions for short distances (up to 2 nm) but reveal a new tail for larger distances in the range of 2.5 – 4 nm (see Local Thickness plot in section 4 of the Supporting Information).

These observations are in general agreement with previous findings by Plank et al.[13], who suggested the formation of larger crystallites as a consequence of continued fragmentation of incompletely or even non-dissociated precursor molecules during EBC. Pt atoms, which are released during that process, then attach at nearest grains, which, in turn, increase their diameter without changing center-to-center distances.[26] The grain shapes, turned out to be more elliptical and grains often interconnect as visible in figure 2. To study this aspect in more detail, we identified interconnected grains, as represented by different colors in figure 4. Surprisingly, such interconnected volumes are already found in as-deposited nano-pillars (a) across the entire volume. (For a better visualization see Video 1 and Video 2 in the Supporting Information.) After EBC treatment (b), similar tendencies are observed, yet for larger individual grain sizes (figure 4). From these findings, we conclude that EBC not only leads to a growth of individual grains (figure 3a) but also to spatial growth of interconnected regions (figure 3b). Due to the size and rigidity of the latter, a homogenous decrease of grain-to-grain distances is not possible any more, explaining tails appearing above 2 nm in the related analysis of figure 3c. Interconnected Pt grain networks lead to non-spherical shapes. Considering previous observations by Huth et al., who proposed that percolation eventually replaces tunneling transport, we can conclude that the here found changes in grain sizes, shapes and mutual distances gets even more dominant for high EBC doses, while the nanogranular character gradually vanishes.[2] This change in morphology can also have strong consequences on mechanical properties, where the metal will play a larger role.

In conclusion, we have applied STEM based 3D nano-tomography to FEBID based Pt-C nano-pillars to reveal the true nature of Pt grain-shapes, -sizes and their spatial distribution in 3D space. While often assumed as close-to-spherical and well separated in the past, our study reveals a much stronger elliptic shape, often the consequence of merged nanograins. Additionally, we have found that even as-deposited materials can contain interconnected grain networks, which increase in volume after post-growth electron beam curing. EBC treatment results in even large interconnected metal networks, which reduces the nano-granular characteristics such as tunneling based electric transport and matrix dominated mechanical properties. Using suitable curing doses, the physical properties of FEBID materials can be tuned from nano-granular towards this interconnected metal network. Especially when high electrical conductivity is needed, EBC with high electron doses can be applied in order to provide such materials due to the formed percolation paths.

FIGURES

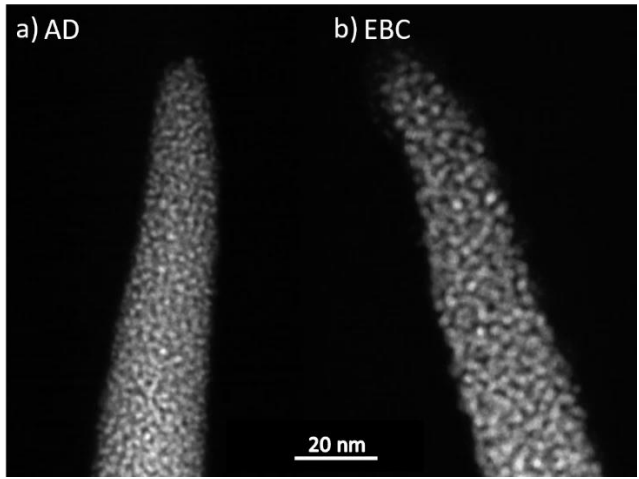

Figure 1. 2D HAADF STEM projection images of Pt-C nano-pillar: (a) after deposition (AD), and (b) after electron-beam curing (EBC). An increase in grain size is visible, comparing AD with EBC.

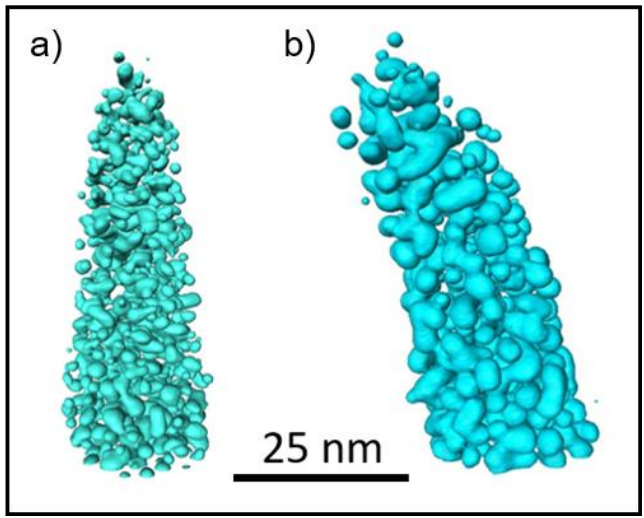

Figure 2. Reconstruction of AD (a) and EBC (b) nano-pillars in the topmost tip region. By a direct comparison the growth of grains with EBC is observable.

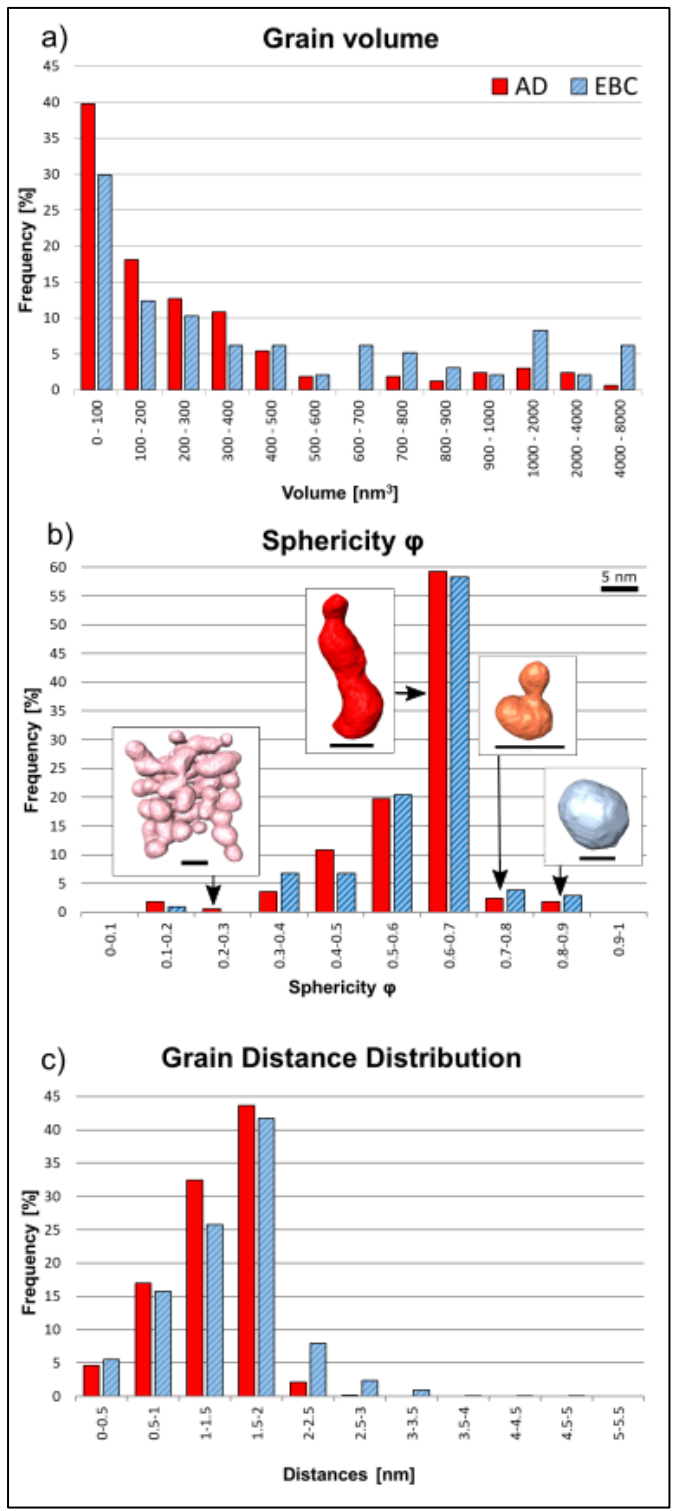

Figure 3. Tomography based analyses of grain volumes (a), -shapes (b) and inter-granular distances (c) for as-deposited (red bars) and EBC nano-pillars (blue bars). As evident, grain sizes

9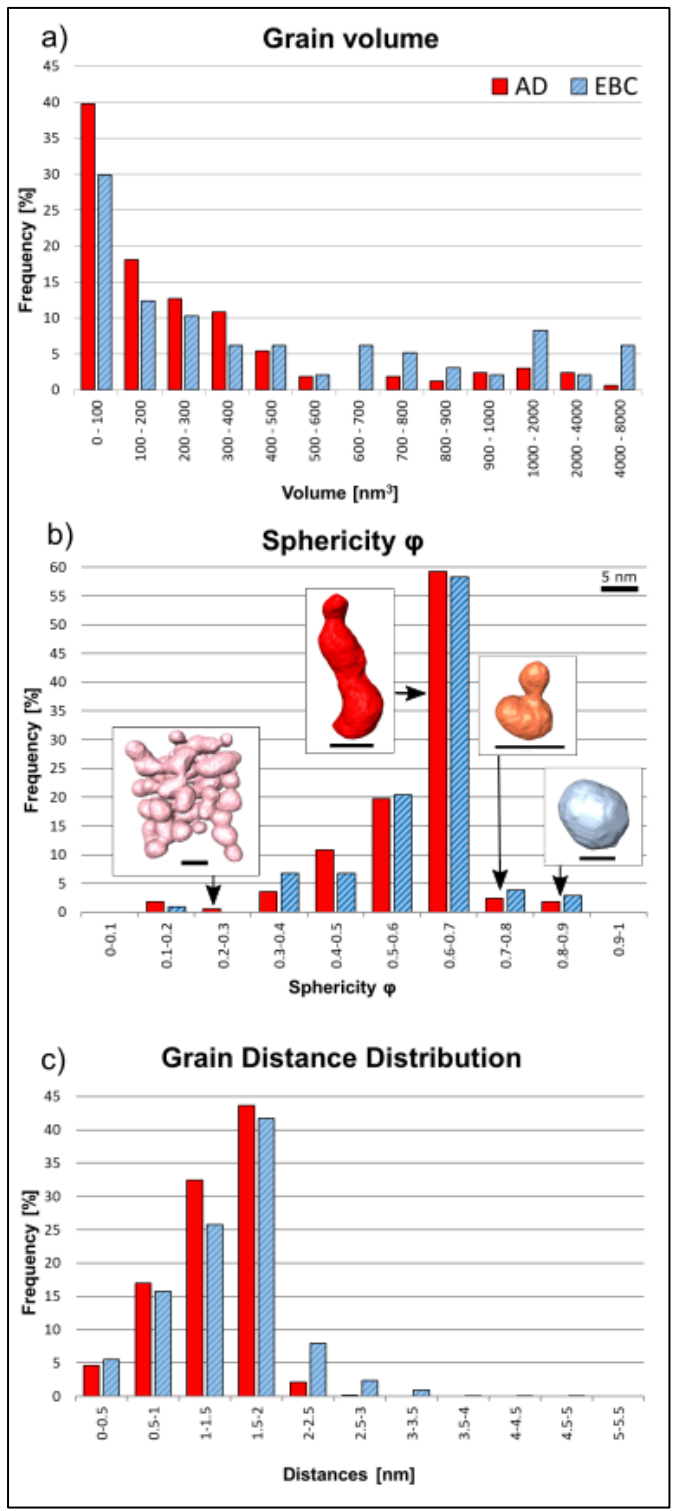

Figure 3. Tomography based analyses of grain volumes (a), -shapes (b) and inter-granular distances (c) for as-deposited (red bars) and EBC nano-pillars (blue bars). As evident, grain sizes

slightly increase after EBC and also reveal to slightly broader size distributions (a). The sphericity φ analysis (b) shows different morphological structures from almost round to very large connected grains. The inter-granular distances of EBC nano-pillars reveal similar behavior up to about 2 nm but also larger distances up to 3.5 nm, in agreement with the coalescence tendencies as described in the main text.

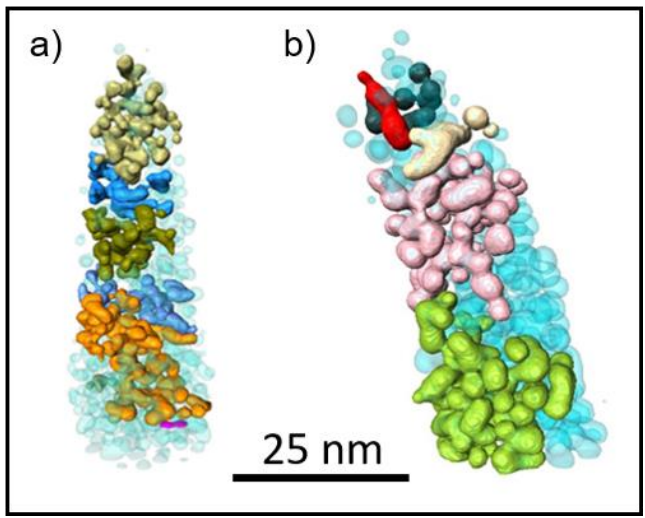

Figure 4. Reconstruction of AD (a) and EBC (b) nano-pillars in the topmost tip region. Connected grains are indicated by the same color and are present in both structures, EBC pillar shows larger volumes of the connected parts.

## ASSOCIATED CONTENT

**Supporting Information.**

Methods including sample preparation, electron energy loss spectroscopy and electron tomography acquisition and processing and local thickness plots. This material is available free of charge via the Internet at http://pubs.acs.org.

## AUTHOR INFORMATION

**Corresponding Author**

*Corresponding authors: georg.haberfehlner@felmi-zfe.at; gerald.kothleitner@felmi-zfe.at

**Author Contributions**

The manuscript was written through contributions of all authors. All authors have given approval to the final version of the manuscript.

## ACKNOWLEDGMENT

RW and HP acknowledge the financial support by the Christian Doppler Research Association (CDL-DEFINE), Austrian Cooperative Research (ACR), FFG Beyond Europe project (AIM, Nr. 11056459). The financial support by the Austrian Federal Ministry for Digital and Economic Affairs and the National Foundation for Research, Technology and Development is gratefully acknowledged. This project has received funding from the European Union's Horizon 2020 research and innovation program under grant agreement No. 823717 – ESTEEM3.